\documentclass{pasj00}   
\usepackage{mathrsfs} 

\begin{document} 

\SetRunningHead{Y. Sofue}{Rotation Curves and Dark Halos}
\Received{} \Accepted{2015} 

\title{Rotation Curve Decomposition for Size-Mass Relations of Bulge, Disk, and Dark Halo in Spiral Galaxies}

\author{Yoshiaki {\sc Sofue} \\ 
Institute of Astronomy, University of Tokyo, Mitaka, 181-0015 Tokyo, Japan \\
Email:{\it sofue@ioa.s.u-tokyo.ac.jp}}
 

\maketitle 

\def\bc{\begin{center}} \def\ec{\end{center}} 
\def\kms{km s$^{-1}$} \def\Msun{M_\odot}
\def\be{ \begin{equation}} \def\ee{\end{equation} }
\def\ab{a_{\rm b}} \def\ad{a_{\rm d}} \def\Mb{M_{\rm b}} \def\Md{M_{\rm d}} 
\def\mb{M_{\rm b}} \def\md{M_{\rm d}} \def\Mh{M_{\rm h}} 
\def\Mbd{M_{\rm b+d}} \def\mbd{M_{\rm b+d}} 
\def\mh{M_{\rm h}} \def\mhalo{M_{200}} \def\mphoto{\mathscr{M}}
\def\rhalo{R_{200}} \def\mbdh{M_{\rm 200+b+d}}
\def\sb{\rm SMD_b} \def\sd{\rm SMD_d} 
 \def\dv{de Vaucouleurs } \def\ha{H$\alpha$}
\def\mtwelve{10^{12}\Msun} \def\meleven{10^{11}\Msun} \def\mten{10^{10}\Msun} 
\def\r{\bibitem[]{}} 
\def\log{{\rm log}_{10}}
\def\rhosp{\rho_{\rm S}}
\def\hub{\mathscr{H}_{72}}
\def\kmsmpc{km s$^{-1}$ Mpc$^{-1}$ } 
\def\bbf{ }

\begin{abstract} 
Rotation curves of more than one hundred spiral galaxies were compiled from the literature, and deconvolved into bulge, disk, and dark halo using $\chi^2$ fitting in order to determine their scale radii and masses. Correlation analyses were obtained of the fitting parameters for galaxies that satisfied selection and accuracy criteria. Size-mass relations indicate that the sizes and masses are positively correlated among different components in such a way that the larger or more massive is the dark halo, the larger or more massive are the disk and bulge. Empirical size-mass relations were obtained for bulge, disk and dark halo by the least-squares fitting. The disk-to-halo mass ratio was found to be systematically greater by a factor of three than that predicted by cosmological simulations combined with photometry. A preliminary mass function for dark halo was obtained, which is represented by the Schechter function followed by a power law. 
\end{abstract}

\section{Introduction}

Decomposition of rotation curves into mass components is an efficient tool to study the dynamical structure of spiral galaxies (Rubin et al. 1980, 1982; Persic and Salucci 1995, 1996; Sofue and Rubin 2001; Noordermeer et al. 2007; de Blok et al. 2008; Martinsson et al. 2013; Sofue 2013, 2015), and is complimentary to photometric decomposition of luminous bulge and disk (Kent 1985; de Jong 1996; Yoshino and Ichikawa 2008; Allen et al. 2006; Bershaldy et al. 2010). 

Rotation curves are particularly useful for measuring dark halos, and hence to derive such fundamental quantities like the mass ratio of the bulge and/or disk to the dark halo. The bulge/disk to dark halo mass ratio is often taken as an indicator for cosmological structure formation and evolution (Reyes et al. 2012; Miller et al. 2014; Moster et al. 2013; Behroozi et al. 2013; and the literature cited therein).  

In rotation curve decomposition it is crucial to employ plausile fitting functions. The \dv or Sersic law spheroid and exponential thin disk models are commonly used for the bulge and disk, respectively. For the dark halo the NFW (Navarro et al 1997, 1997) model was shown to be most plausible to represent the recent precise rotation curves up to several hundred kpc of the Galaxy and M31, whereas the isothermal sphere model cout not fit the observations (Sofue 2013, 2015). 

In the present paper, rotation curves are compiled from the literature in the last two decades, and are deconvolved into the three mass components using the $\chi^2$ method described in Sofue (2013, 2015). The fitted parameters will be used to derive various relations among sizes and masses of bulges, disks, and dark halos.  
 
\section{Compilation of Rotation Curves and Decomposition Method} 
 
\begin{figure*} 
\begin{center}
\includegraphics[width=15cm ]{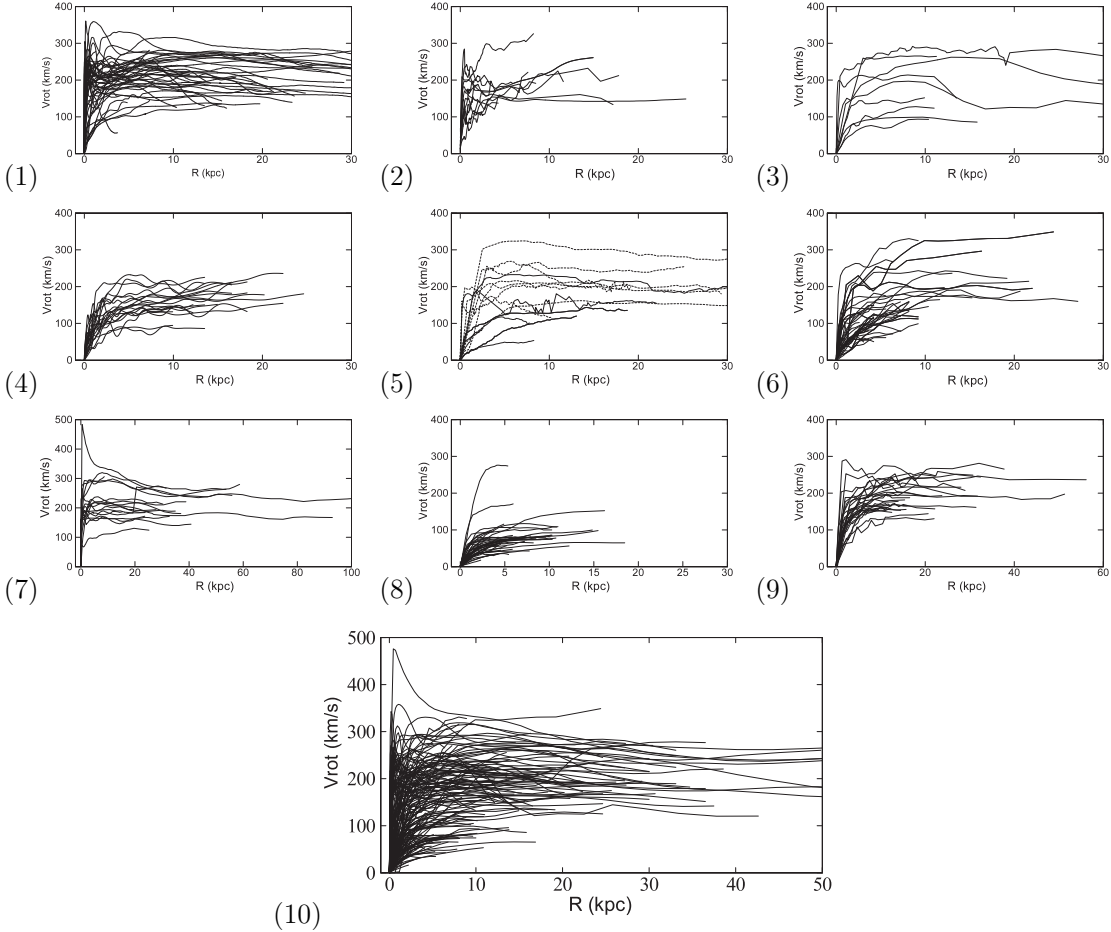} 
\end{center}
\caption{Rotation curves compiled and reproduced from the literature. References are in the order of panel numbers,
(1) Sofue et al. (1999): Nearby galaxy rotation curve atlas; 
(2) Sofue et al. (2003): Virgo galaxy CO line survey; 
(3) Sofue et al. (1999: NGC 253 revised); Ryder et al. (1998, NGC157); Hlavacek-Larrondo et al. (2011a: NGC253; 2011b: NGC 300); Erroz-Ferrer et al. (2012: NGC 864); Gentile et al. (2015: NGC 3223); Olling R.~P. (1996: NGC 4244); Whitmore \& Schweizer (1987: NGC 4650A); 
Gentile et al. (2007: NGC 6907); 
(4) Marquez et al. (2004): Isolated galaxy survey;
(5) de Blok et al. (2008): THINGS survey, where dashed galaxies are included in (1) and were not used in the analysis; 
(6) Garrido et al. (2005): GHASP survey; 
(7) Noordermeer et al. (2007): Early type spiral survey; 
(8) Swaters et al. (2009): Dwarf and low-surface-brightness galaxy survey; 
(9) Martinsson et al. (2013): DiskMass survey;
and
(10) All rotation curves in one panel.} 
\label{allrc} 
\end{figure*}

Rotation curves were compiled from the observational data from 
Sofue et al. (1999); 
Sofue et al. (2003);
 M{\'a}rquez et al. (2004):
 de Blok et al. (2008); 
 Garrido et al. (2005); 
 Noordermeer et al. (2007); 
 Swaters et al. (2009); and
Martinsson et al. (2013),
and are shown in figure \ref{allrc}. 
Data for individual galaxies were also compiled from the literature and are shown in the caption to the figure. Data without digitized presentation were read from the published figures using a graph-reading tool. The accuracy of the reading was about $\pm 0.2$ kpc in radius and $\pm 3$ \kms in velocity, which were sufficiently smaller than the fitting accuracy and the dispersion among the data. Sample galaxies were so selected that they had end radii of observed rotation curves greater than 10 kpc and the number of data points in a curve was sufficiently larger compared to the number of fitting parameters. All used rotation curves are presented in Appendix 1.

Rotation curve, $V(R)$, is composed of the following three components expressed by
\be
V(R)^2=V_{\rm b}(R)^2+V_{\rm d}(R)^2+V_{\rm h}(R)^2,
\label{vrot}
\ee 
where $V_{\rm b}(R),~ V_{\rm d}(R)$, and $V_{\rm h}(R)$ are the rotation velocity at galacto-centric distance $R$ corresponding to the bulge, disk and dark halo, respectively. 
 
The bulge is assumed to have the \dv (1958) profile for the surface mass density as 
\be 
\Sigma_{\rm b}(R) = \Sigma_{\rm be} {\rm exp} 
\left[-\kappa \left\{\left(R / a_{\rm b} \right)^{1/4}-1\right\}\right],
\ee 
where $\kappa=7.6695$, $\Sigma_{\rm be}$ is the surface mass density at the half-surface mass scale radius $R=\ab$. The total mass of the bulge, $\Mb$, is calculated by $\Mb= 22.665 \ab^2 \Sigma_{\rm be}$ using the scale radius $\ab$. In the fitting, $\ab$ and $\Mb$ were taken as the two free parameters. Sersic model with n=2 would be an alternative choice. However, the major goal of the present paper is studying halos, and the central rotation curves have no sufficient resolution to discriminate the Sersic indices. Hence, the classical value of $n=4$ is adopted here.
 
The galactic disk is approximated by an exponential thin disk (Freeman 1970), as inferred from surface photometry.
This holds even if the interstellar gas (molecular and neutral hydrogen) is included, that shares $\sim 10$\% of the disk mass (Nakanishi and Sofue 2006). 
Then, the surface mass density is expressed by
\be 
\Sigma_{\rm d} (R)=\Sigma_0 {\rm exp} \left( -{R / \ad } \right), 
\label{disksigma}
\ee 
where $\Sigma_0$ is the central value and $\ad$ is the scale radius. In the fitting, $\ad$ and $\Sigma_0$ were taken as the two free parameters. The results are presented in terms of $\ad$ and the total disk mass $\Md=2 \pi \ad^2 \Sigma_0$. 

For the dark halo, the NFW (Navarro et al. 1996, 1977) density profile is assumed, which was shown to be a reasonable model in the outer dark halos of the Milky Way and M31 up to radii as large as $\sim 400$ kpc (Sofue 2013, 2015). The NFW profile is expressed as 
\be 
\rho(R)={\rho_0 /[ X\left(1+ X \right)^2] } ,
\label{nfw}
\ee 
where $X={R/ h}$, and $\rho_0$ and $h$ are the representative density and scale radius of the dark halo, respectively. In the fitting procedure, $\rho_0$ and $h$ were taken as the two free parameters. The enclosed mass within radius $R$ is given by
\be 
\Mh (R) 
 = 4 \pi \rho_0 h^3 \left\{ {\rm ln} (1+X)-{X /( 1+X)}\right\}.
\label{mh} 
\ee
The rotation velocity is given by
\be
V_{\rm h} (R)=\sqrt{G \Mh (R) / R} .
\label{vh}
\ee 

The critical mass $M_{200}$ and radius $R_{200}$ (Navarro et al. 1997) are defined as the following, where $\rho_{\rm c}=3H^2_0/8\pi G$ with $H_0 =72 $ \kmsmpc (Hinshaw et al. 2009) being the Hubble constant and $G$ the gravitational constant:
\be
M_{200}=200 \rho_{\rm c}{4\pi \over 3} R_{200}^3.
\label{m200}
\ee 
Note that the calculated mass is dependent on the Hubble constant as $\mhalo\propto H_0^2$, and the mass may be multiplied by ${\hub} ^2$ to convert to a value for a different $H_0$ with $\hub=H_0/72$ being the correctio factor.
Defining $X_{200}$ by
\be
X_{200}=R_{200}/h,
\label{x200r200}
\ee
and using equation (\ref{mh}), it is shown that
\be 
{\rm ln} (1+X_{200})-{X_{200} /( 1+X_{200})}={200 \rho_{\rm c}\over 3\rho_0} {X_{200}^3}.
\label{x200rho0} 
\ee 
 Given a set of parameters $\rho_0$ and $h$ by the rotation curve fitting, the characteristic mass and radius $M_{200}$ and $R_{200}$ are calculated using equations (\ref{m200}) and (\ref{x200r200}), solving equation (\ref{x200rho0}) by succesive approximation.

The free fitting parameters were $\Sigma_{\rm be},~ \ab,~ \Sigma_0,~ \ad,~ \rho_0$ and $h$. The masses of the bulge $\Mb$ and disk $\Md$, the critical dark halo radius $R_{200}$, critical mass $M_{200}$, as well as the halo mass $M_h$ enclosed within $h$ were also calculated. The results are presented in terms of $\ab,~\Mb,~ \ad,~ \Md,~ h, \mh,~ \rhalo,$ and $\mhalo$.

Search for the best fit was made iteratively from one pair of parameters after another among the three pairs representing bulge, disk, and dark halo (Sofue 2013). Each search was made in two dimensional parameter space of each pair for bulge, disk or halo, but not in the entire six-dimensional space. Each $\chi^2$ minimum represents, therefore, the best fit for a parameter pair with the other two pairs fixed to their former best values.  Fitting radii were taken to be 0 to Min[$r_{\rm max}, R_{\rm max}$], where $r_{\rm max}$ is the end radius of observation, and $R_{\rm max}$ is the maximum radius of fitting with $R_{\rm max}=10,$ 20 and 100 kpc for bulge, disk and halo, respectively.

The fitting was obtained using the modified $\chi^2$ fitting as described in Sofue (2013, 2015), where the errors in the observed velocities were taken to be unity. The uncertainties of the fitted parameters were estimated as the ranges of parameter values around the $\chi^2$ minimum that allow for 10\% greater $\chi^2$ values above the least values. 
{\bbf 
The simplified method was employed, because some of the compiled data had no error indication and some data were regridded. In order to confirm that this simplification did not cause artificially better fitting, the usual $\chi^2$ method was applied by assuming a constant dispersion of 10 \kms as a typical error for the observed velocities. Thus computed uncertainties in fitted parameters in terms of the curvature of the $\chi^2$ function near the minimum were shown to be smaller than those by the simplified method.
}

\begin{figure*}
\begin{center}
\includegraphics[width=12cm]{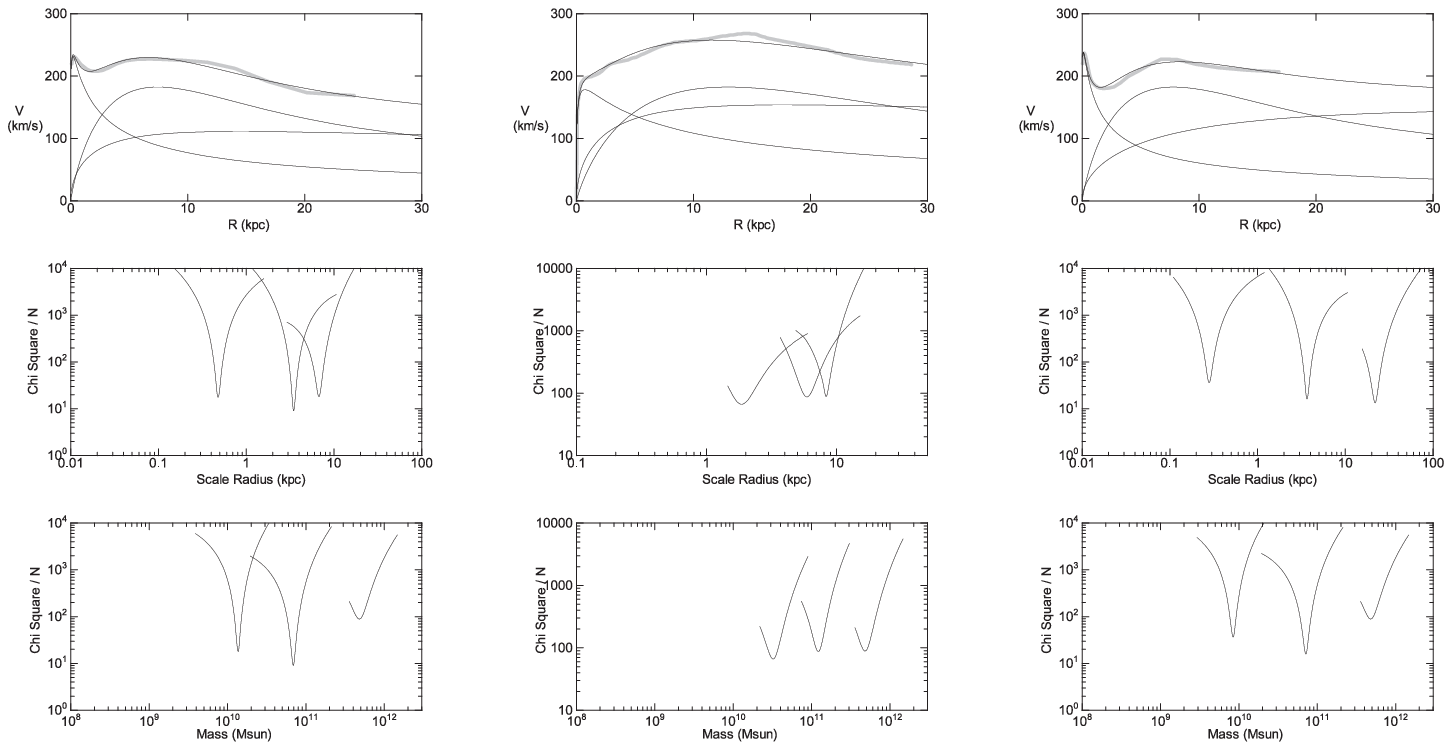}
\end{center}
\caption{[Top panels] Examples of rotation curves and fitting results for NGC 891 (left), NGC 5907 (middle) and NGC 6946 (right). 
[2nd] Distributions of $\chi^2/N$ around the best fit scale radii. 
[bottom] Same, but around the best fit masses. 
} 
\label{rc} 
\end{figure*}

Figure \ref{rc} shows examples of rotation curves and fitting results for NGC 891, NGC 5907 and NGC 6946. The figure also shows the variation of $\chi^2/N$ values plotted against the parameters.

The rotation curve decomposition was applied to all the compiled galaxies, and displayed in Appendix. However, many of the resulting parameters were not necessarily reasonable to represent realistic galactic structures, and were not accurate enough for quantitative statistical analyses. We, therefore, rejected galaxies that showed scale radii of bulge and disk anomalously greater than those of disk and halo, respectively, in each galaxy. Namely, galaxies were so chosen that they satisfied the condition in each galaxy,
\be
\ab<2\ad<4h.
\label{condition}
\ee
Galaxies which had anomalously large halo radii with $h>200$ kpc were rejected, which was too far beyond the observed radii. Also, galaxies, one of whose parameters had a larger error than the parameter value itself, were rejected. Namely, another condition was added so that
\be
\delta p_i<p_i,
\ee
where $\delta p_i$ is the error of parameter $p_i$.

\section{Results}

Applying the above condition, 43 galaxies were selected among the analyzed galaxies. The best-fitting parameters are listed in table \ref{bdhpara}. It should be remembered that, since the derived masses are dynamical masses from rotation curve decomposition, the disk and bulge masses might contain dark matter. Also, the dark halo mass might contain baryonic mass. 
 
\renewcommand{\arraystretch}{0.75} 
\begin{table*} 
\bc
\caption{Dynamical parameters for selected spiral galaxies.} 
\tabcolsep3pt  
\footnotesize  
\begin{tabular}{lllllllllllll}
\hline\hline \\
 Name$^\dagger$ & $r_{\rm max}^\ddagger$ & $\ab$(kpc)& $\mb(\mten)$& $\ad$(kpc) & $\md(\mten)$ & $h$(kpc) & $M_{\rm h}(\mten)$ &$\rhalo$(kpc) & $\mhalo(\mten)^\ddagger$ \\
\hline 
 \\
 100000& 469.6& 1.52$\pm$ 0.11& 3.53$\pm$ 0.33& 5.9$\pm$ 0.1& 9.2$\pm$ 0.4& 14.7$\pm$ 0.4& 9.2$\pm$ 0.6& 190.$\pm$ 11.& 81.$\pm$ 7.\\
 100224& 438.9& 1.30$\pm$ 0.06& 3.36$\pm$ 0.16& 4.3$\pm$ 0.3& 7.7$\pm$ 0.8& 30.5$\pm$ 0.7& 27.9$\pm$ 1.3& 253.$\pm$ 10.& 193.$\pm$ 15.\\
 100342& 19.3& 0.64$\pm$ 0.06& 0.52$\pm$ 0.04& 1.6$\pm$ 0.1& 1.6$\pm$ 0.2& 12.0$\pm$ 0.3& 7.9$\pm$ 0.4& 185.$\pm$ 8.& 76.$\pm$ 6.\\
 100598& 15.4& 7.66$\pm$ 2.94& 0.22$\pm$ 0.11& 6.1$\pm$ 2.1& 0.5$\pm$ 0.3& 8.5$\pm$ 0.2& 2.4$\pm$ 0.1& 124.$\pm$ 5.& 23.$\pm$ 2.\\
 100660& 23.3& 0.57$\pm$ 0.05& 0.94$\pm$ 0.07& 0.6$\pm$ 0.2& 0.2$\pm$ 0.1& 9.2$\pm$ 0.2& 3.8$\pm$ 0.2& 147.$\pm$ 6.& 37.$\pm$ 3.\\
 \\
 100891& 24.8& 0.71$\pm$ 0.05& 1.84$\pm$ 0.13& 3.1$\pm$ 0.4& 3.1$\pm$ 0.5& 6.4$\pm$ 0.2& 3.1$\pm$ 0.3& 144.$\pm$ 11.& 35.$\pm$ 4.\\
 101365& 31.1& 0.90$\pm$ 0.09& 2.51$\pm$ 0.18& 3.3$\pm$ 0.2& 6.6$\pm$ 0.8& 14.4$\pm$ 0.3& 10.4$\pm$ 0.9& 199.$\pm$ 16.& 95.$\pm$ 10.\\
 101808& 16.0& 0.66$\pm$ 0.06& 1.38$\pm$ 0.10& 3.0$\pm$ 0.3& 3.2$\pm$ 0.4& 2.0$\pm$ 0.0& 0.4$\pm$ 0.0& 77.$\pm$ 8.& 6.$\pm$ 1.\\
 102403& 19.7& 0.14$\pm$ 0.05& 0.02$\pm$ 0.00& 0.2$\pm$ 0.0& 0.0$\pm$ 0.0& 7.6$\pm$ 0.2& 2.9$\pm$ 0.1& 135.$\pm$ 6.& 29.$\pm$ 2.\\
 102903& 23.8& 2.50$\pm$ 0.30& 5.24$\pm$ 0.63& 3.8$\pm$ 0.5& 8.0$\pm$ 1.3& 7.6$\pm$ 0.5& 5.0$\pm$ 1.0& 168.$\pm$ 33.& 56.$\pm$ 16.\\
 \\
 103079& 21.3& 0.69$\pm$ 0.07& 2.63$\pm$ 0.19& 3.8$\pm$ 0.6& 4.4$\pm$ 0.8& 16.2$\pm$ 1.2& 9.5$\pm$ 1.6& 189.$\pm$ 30.& 80.$\pm$ 21.\\
 103198& 31.1& 6.21$\pm$ 1.46& 0.50$\pm$ 0.10& 3.1$\pm$ 0.2& 1.6$\pm$ 0.2& 13.4$\pm$ 0.3& 6.2$\pm$ 0.5& 166.$\pm$ 13.& 54.$\pm$ 6.\\
 103628& 14.2& 0.84$\pm$ 0.08& 1.52$\pm$ 0.11& 3.6$\pm$ 0.4& 3.6$\pm$ 0.5& 8.5$\pm$ 0.2& 4.8$\pm$ 0.4& 162.$\pm$ 13.& 51.$\pm$ 5.\\
 104258& 29.2& 0.47$\pm$ 0.04& 1.19$\pm$ 0.09& 0.8$\pm$ 0.1& 0.9$\pm$ 0.1& 19.4$\pm$ 0.5& 16.1$\pm$ 1.0& 225.$\pm$ 13.& 134.$\pm$ 12.\\
 104321& 25.6& 1.32$\pm$ 0.09& 2.76$\pm$ 0.20& 7.6$\pm$ 0.5& 16.8$\pm$ 1.6& 7.1$\pm$ 0.2& 4.2$\pm$ 0.3& 159.$\pm$ 13.& 48.$\pm$ 5.\\
 \\
 104527& 12.8& 0.41$\pm$ 0.08& 0.94$\pm$ 0.13& 2.6$\pm$ 0.8& 1.7$\pm$ 0.6& 12.2$\pm$ 1.5& 7.4$\pm$ 2.0& 181.$\pm$ 43.& 70.$\pm$ 29.\\
 104565& 34.1& 3.05$\pm$ 0.28& 6.40$\pm$ 0.76& 4.0$\pm$ 0.8& 4.6$\pm$ 1.2& 19.2$\pm$ 1.4& 16.3$\pm$ 2.8& 226.$\pm$ 35.& 137.$\pm$ 35.\\
 104736& 10.4& 0.82$\pm$ 0.12& 1.09$\pm$ 0.13& 0.9$\pm$ 0.2& 0.8$\pm$ 0.2& 7.3$\pm$ 0.7& 2.3$\pm$ 0.5& 125.$\pm$ 27.& 23.$\pm$ 8.\\
 104945& 20.0& 0.36$\pm$ 0.04& 0.69$\pm$ 0.07& 0.3$\pm$ 0.1& 0.3$\pm$ 0.0& 9.0$\pm$ 0.2& 5.6$\pm$ 0.3& 170.$\pm$ 10.& 59.$\pm$ 5.\\
 105033& 35.1& 1.04$\pm$ 0.10& 3.76$\pm$ 0.27& 5.7$\pm$ 0.7& 11.4$\pm$ 1.6& 38.9$\pm$ 2.8& 41.2$\pm$ 7.1& 280.$\pm$ 44.& 262.$\pm$ 68.\\
 \\
 105055& 39.4& 2.96$\pm$ 0.21& 4.15$\pm$ 0.30& 1.8$\pm$ 0.3& 1.7$\pm$ 0.2& 7.7$\pm$ 0.2& 4.2$\pm$ 0.3& 156.$\pm$ 9.& 45.$\pm$ 4.\\
 105236& 39.3& 0.19$\pm$ 0.04& 0.47$\pm$ 0.06& 3.0$\pm$ 0.7& 1.8$\pm$ 0.5& 8.1$\pm$ 0.2& 4.4$\pm$ 0.4& 158.$\pm$ 12.& 47.$\pm$ 5.\\
 105457& 13.5& 2.76$\pm$ 0.41& 2.90$\pm$ 0.41& 2.4$\pm$ 0.5& 2.2$\pm$ 0.5& 8.1$\pm$ 0.6& 4.6$\pm$ 0.9& 161.$\pm$ 28.& 50.$\pm$ 13.\\
 105907& 28.6& 1.60$\pm$ 0.08& 2.76$\pm$ 0.13& 6.9$\pm$ 0.5& 13.8$\pm$ 1.0& 6.7$\pm$ 0.2& 3.4$\pm$ 0.2& 148.$\pm$ 6.& 39.$\pm$ 3.\\
 106946& 17.0& 0.36$\pm$ 0.04& 0.92$\pm$ 0.07& 3.8$\pm$ 0.5& 4.2$\pm$ 0.6& 9.4$\pm$ 0.2& 5.6$\pm$ 0.5& 169.$\pm$ 13.& 58.$\pm$ 6.\\
 \\
 103521& 23.6& 0.72$\pm$ 0.09& 1.52$\pm$ 0.15& 1.6$\pm$ 0.2& 3.5$\pm$ 0.3& 22.6$\pm$ 1.6& 14.0$\pm$ 2.4& 206.$\pm$ 32.& 103.$\pm$ 26.\\
 204303& 25.3& 0.33$\pm$ 0.14& 0.15$\pm$ 0.05& 2.1$\pm$ 0.5& 1.1$\pm$ 0.3& 11.9$\pm$ 0.9& 4.5$\pm$ 0.8& 148.$\pm$ 26.& 39.$\pm$ 10.\\
 204569& 14.9& 1.39$\pm$ 0.37& 0.66$\pm$ 0.18& 12.0$\pm$ 0.9& 39.7$\pm$ 5.8& 9.8$\pm$ 1.6& 3.5$\pm$ 1.3& 140.$\pm$ 47.& 33.$\pm$ 19.\\
 300253& 14.2& 0.93$\pm$ 0.09& 1.60$\pm$ 0.11& 1.9$\pm$ 0.3& 1.9$\pm$ 0.4& 7.9$\pm$ 0.6& 3.5$\pm$ 0.6& 145.$\pm$ 23.& 36.$\pm$ 9.\\
 600508& 25.7& 2.89$\pm$ 1.31& 2.12$\pm$ 0.84& 6.6$\pm$ 0.5& 25.1$\pm$ 3.0& 87.1$\pm$ 10.4& 239.3$\pm$ 59.7& 479.$\pm$ 105.& 1271.$\pm$ 515.\\
 \\
 604273& 13.0& 0.99$\pm$ 0.94& 0.15$\pm$ 0.08& 3.8$\pm$ 0.3& 4.0$\pm$ 0.5& 20.6$\pm$ 1.5& 14.1$\pm$ 2.2& 210.$\pm$ 29.& 110.$\pm$ 27.\\
 702916& 36.9& 1.56$\pm$ 0.22& 2.18$\pm$ 0.26& 2.8$\pm$ 0.6& 2.6$\pm$ 0.5& 11.4$\pm$ 0.3& 7.5$\pm$ 0.6& 183.$\pm$ 14.& 73.$\pm$ 8.\\
 703993& 53.6& 4.35$\pm$ 0.31& 16.59$\pm$ 1.19& 4.8$\pm$ 0.7& 7.5$\pm$ 1.3& 135.1$\pm$ 3.2& 214.9$\pm$ 17.7& 392.$\pm$ 31.& 687.$\pm$ 83.\\
 704458& 55.9& 1.08$\pm$ 0.15& 10.03$\pm$ 1.20& 3.5$\pm$ 0.6& 12.5$\pm$ 2.1& 50.2$\pm$ 6.0& 43.8$\pm$ 11.6& 271.$\pm$ 64.& 229.$\pm$ 97.\\
 704605& 34.9& 1.07$\pm$ 0.10& 1.45$\pm$ 0.10& 2.8$\pm$ 0.1& 5.7$\pm$ 0.4& 52.9$\pm$ 3.8& 38.2$\pm$ 5.5& 249.$\pm$ 31.& 181.$\pm$ 43.\\
 \\
 705253& 40.7& 0.56$\pm$ 0.14& 0.39$\pm$ 0.07& 1.1$\pm$ 0.2& 1.1$\pm$ 0.1& 9.7$\pm$ 0.2& 4.2$\pm$ 0.3& 151.$\pm$ 9.& 41.$\pm$ 4.\\
 706786& 30.3& 1.66$\pm$ 0.12& 2.44$\pm$ 0.17& 6.3$\pm$ 0.3& 11.9$\pm$ 0.9& 7.7$\pm$ 0.4& 2.8$\pm$ 0.3& 133.$\pm$ 13.& 28.$\pm$ 5.\\
 709113& 102.3& 0.83$\pm$ 0.14& 1.11$\pm$ 0.13& 0.7$\pm$ 0.3& 0.5$\pm$ 0.2& 25.1$\pm$ 0.6& 29.8$\pm$ 1.8& 273.$\pm$ 16.& 241.$\pm$ 22.\\
 711670& 23.6& 0.82$\pm$ 0.14& 0.64$\pm$ 0.11& 0.8$\pm$ 0.1& 0.8$\pm$ 0.1& 11.8$\pm$ 0.8& 5.7$\pm$ 0.8& 164.$\pm$ 20.& 52.$\pm$ 12.\\
 711852& 93.0& 0.61$\pm$ 0.10& 0.94$\pm$ 0.11& 3.3$\pm$ 0.2& 6.3$\pm$ 0.4& 29.5$\pm$ 0.7& 15.1$\pm$ 0.7& 197.$\pm$ 8.& 92.$\pm$ 7.\\
 \\
 903140& 31.8& 3.06$\pm$ 0.66& 2.78$\pm$ 0.47& 3.0$\pm$ 0.2& 6.0$\pm$ 0.4& 88.9$\pm$ 2.1& 83.1$\pm$ 6.9& 293.$\pm$ 23.& 297.$\pm$ 36.\\
 904555& 16.5& 1.56$\pm$ 0.94& 0.16$\pm$ 0.08& 3.1$\pm$ 0.2& 3.1$\pm$ 0.1& 19.7$\pm$ 0.5& 11.7$\pm$ 0.6& 197.$\pm$ 8.& 90.$\pm$ 7.\\
 906918& 12.2& 0.55$\pm$ 0.21& 0.09$\pm$ 0.03& 1.2$\pm$ 0.1& 1.2$\pm$ 0.1& 23.6$\pm$ 0.6& 13.8$\pm$ 0.7& 201.$\pm$ 8.& 97.$\pm$ 8.\\
\\
\hline
\end{tabular} 
\ec

$\dagger$ Galaxy names are so given as NGC, IC and UGC numbers added by six digit numbers as :
100000: Sofue (1999) by NGC except for the Galaxy as 0000 and IC342 as 0342 from nearby galaxy compilation;
200000: Sofue (2003) by NGC from Virgo galaxy survey;
300000: Individual galaxies added in this paper.
600000: Garrido et al. (2005) by UGC from GHASP survey;
700000: Noordermeer et al. (2007) by UGC from early type galaxies survey;
800000: Swaters et al. (2009) by UGC from dwarf survey, but were all rejected by the accuracy criterion;
900000: Martinsson et al. (2013) by UGC from DiskMass survey;
500000: de Blok et al. (2008) by NGC from THINGS survey. 

$\ddagger$ Dark halo mass was calculated for $H_0=72$ \kmsmpc, and may be multiplied by a correction factor $\hub^2$ with $\hub=H_0/72$ to obtain a value corresponding to different $H_0$. 

$\ddagger$ End radius of observation. Fitting radii were from 0 to $R_{\rm max}=10,$ 20 and 100 kpc for bulge, disk and halo, respectively.

\label{bdhpara} 
\end{table*} 
 \renewcommand{\arraystretch}{1} 
 
\subsection{Mean parameters} 

Mean values and errors of the fitted parameters for the selected 43 galaxies are listed in table \ref{meanpara}. The ratio of the mean bulge+disk mass to the dark halo critical mass was obtained to be 0.062 $\pm$ 0.018, and that to the total mass 0.059$\pm$0.016. This represents the baryonic fraction within the critical radius. 
  
 The mean baryonic fraction is comparable to the Local Group value for the Galaxy and M31 of $\sim 0.07$ (Sofue 2015). However, it is significantly smaller than the cosmological value of 0.17 from the WMAP (Wilkinson Microwave Anisotropy Probe) observations (e.g., Dunkley et al. 2009). This implies that the galaxies analyzed here are more dark mater dominant compared to the cosmological value. Alternatively, the rest of baryons with mass of $\sim 0.06/0.17 \mhalo \sim 3.5\times 10^{11}\Msun$ might be distributed in the dark halos of radii $\sim 200$ kpc. 

\def\amh{M_{\rm h}}
\begin{table*} 
\caption{Mean parameters for selected galaxies$^\dagger$.}
\bc
\begin{tabular}{lll }
\hline\hline 
 Number of galaxies \dotfill& $N$ & 43\\ 
\\
 Bulge scale radius \dotfill&$\ab$ (kpc)& 1.5$\pm$ 0.2\\
 --- mass \dotfill& $\mb(\mten)$& 2.3$\pm$ 0.4\\ 
 \\
 Disk scale radius \dotfill& $\ad$(kpc)& 3.3$\pm$ 0.3\\
 --- mass \dotfill& $\md(\mten)$& 5.7$\pm$ 1.1\\ 
\\
 Dark Halo scale radius \dotfill& $h$ (kpc)& 21.6$\pm$ 3.9\\
 --- mass within $h$ \dotfill& $\mh(\mten)$& 22.3$\pm$ 7.3\\
 --- critical radius \dotfill& $\rhalo$ (kpc)& 193.7$\pm$ 10.8\\
 --- critical mass \dotfill& $\mhalo (\mten)$ & 127.6$\pm$ 32.0\\ 
\\
 B+D mass\dotfill& $M_{\rm b+d} (\mten)$ & 7.9$\pm$ 1.2\\ 
 B+D+H mass\dotfill& $M_{\rm 200+b+d} (\mten)$ & 135.6$\pm$ 32.0\\
 \\
 (B+D)/Halo mass ratio\dotfill & $M_{\rm b+d}/M_{\rm 200}$ &0.062$\pm$0.018\\ 
 (B+D)/Total mass ratio\dotfill &$M_{\rm b+d}/M_{\rm 200+b+d}$& 0.059$\pm$ 0.016\\

 \hline
\end{tabular} 

$^\dagger$ The uncertainties are standard errors.
\ec
\label{meanpara} 
\end{table*}

\subsection{Size-Size relation}

\begin{figure} 
\bc
\includegraphics[width=75mm]{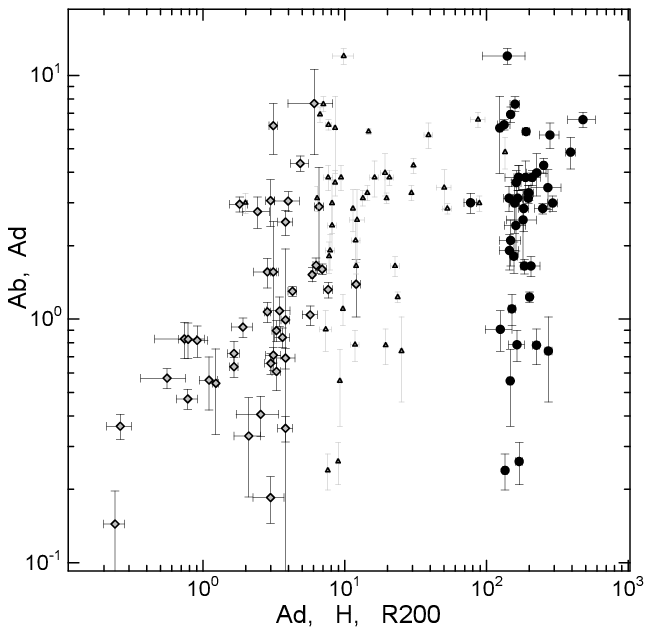} 
\ec
\caption{Size-Size relations for $(x,y)=(\ad,\ab)$ (gray diamonds); $(h,\ad)$ ( gray dot ), and $(\rhalo,\ad)$ (black dots). } 
\label{sizerelation}

\bc
\includegraphics[width=75mm]{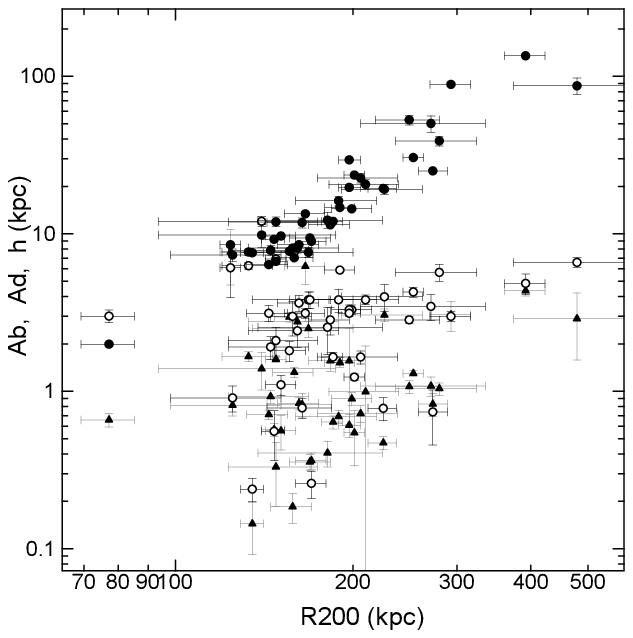} 
\ec
\caption{
Size-Size relations for $(\rhalo,\ab)$ (triangles); $(\rhalo,\ad)$ (open circles); and $(\rhalo,h)$ (black dots). } 
\label{sizerelation_b} 
\end{figure}

The correlations among the parameters were analyzed by plotting them in various parameter spaces. The size-size relations, or plots of the scale radii among the bulge, disk and dark halo, are shown in figure \ref{sizerelation} and \ref{sizerelation_b}. The figures show that the larger is the disk size, the larger is the bulge, and that the larger is the halo size, the larger is the disk. 
 
In order to quantify the significance of correlation, the Spearman's rank correlation coefficient, $\rhosp$, the linear correlation coefficient in a log-log plot, $r$, and the probability for the coefficient exceeding $r$ from uncorrelated sample, $P(r;N)$, were calculated for all the pairs of derived parameters, and are listed in table \ref{spearman}. The coefficients for the size-size relations are all positive, confirming the apparent positive correlation. The size-size rank coefficient between bulge and disk is as high as 0.56.  

It is also noticeable that the logarithmic scatter of critical radius $\rhalo$ in figure \ref{sizerelation} is smaller than those for other components. This reflects the fact that the mean radius of dark halos has a small standard error, $194\pm 11$ (table \ref{meanpara}), corresponding to a standard deviation of $\pm 71$ kpc. The narrow radius range might represent some universal constant about the dark halo size. 

Figure \ref{sizerelation_b} shows plots of $\ab,~ \ad,$ and $h$ against the critical radius $\rhalo$, which shows that the bulge, disk and halo scale radii are positively correlated with $\rhalo$.
Note, however, that the tight correlation between $h$ and $R_{200}$ partially includes a trivial internal relation due to the definition of the two parameters connected by $\rho_0$ through equations (\ref{mh}) and (\ref{m200}).

\begin{table} 
\caption{Spearman's rank correlation coefficient $\rho_{\rm S}$, linear correlation coefficient $r$ for log-log plots, and probability $P(r;N)$ for the coefficient exceeding $r$ from uncorrelated sample. The number of sample galaxies is $N=43$.}
\bc
\begin{tabular}{lllll }
\hline\hline 
param 1 & param 2& $\rho_{\rm S}$ & $r$ & $P(r;N)$\\
\hline 
 $\ab$ & $\ad$ & 0.56& 0.54& $0.0002$\\
 $\ab$ & $h$ & 0.16& 0.28& 0.07 \\
 $\ab$ &$\rhalo$ & 0.19& 0.26& 0.09 \\
 $\ad$ & $h$ & 0.08& 0.16& 0.31 \\
 $\ad$ &$\rhalo$ & 0.11& 0.16& 0.31 \\
 $h$ &$\rhalo$ & 0.90& 0.94& 0\\
\\
 $\mb$ & $\md$ & 0.63& 0.62& $6\times 10^{-5}$\\
 $\mb$ & $\mh$ & 0.29& 0.34& 0.03\\
 $\mb$ & $\mhalo$ & 0.34& 0.34& 0.03\\
 $\md$ & $\mh$ & 0.32 & 0.35 & 0.02\\
 $\md$ & $\mhalo$ & 0.33 & 0.34& 0.03\\
 $\mh$ & $\mhalo$ & 0.99& 0.98 &0\\ 
\\
 $\ab$ & $\mb$ & 0.55 & 0.47& $0.001$\\
 $\ad$ & $\md$ & 0.84 & 0.87& 0\\
 $h$ & $\mh$ & 0.94 & 0.98& 0\\
 $\rhalo$ & $\mhalo$ & 1.00&1.00 & 0\\ 

 \hline
\end{tabular} 
\ec
\label{spearman} 
\end{table}

\subsection{Mass-Mass relation}

\begin{figure} \bc
 \includegraphics[width=75mm]{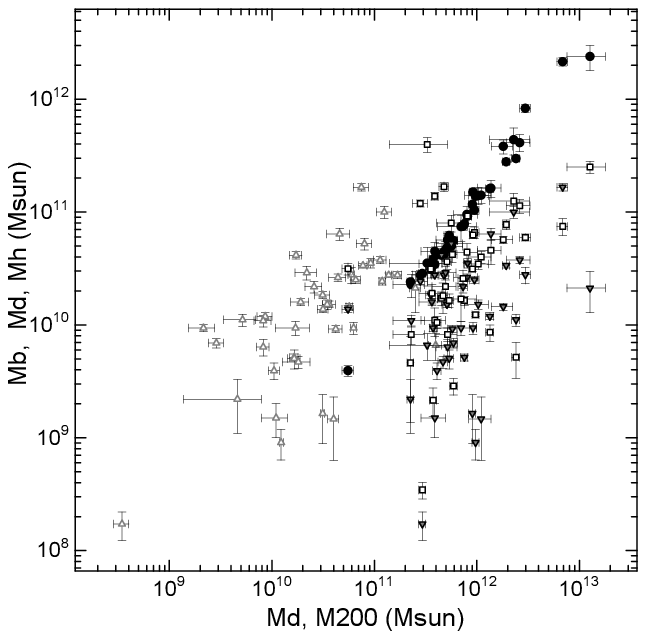} 
 \ec
\caption{Mass-Mass relations for $(x,y)$ $=(\md,\mb)$ (gray triangles); $(\mhalo,\mb)$ (reverse gray triangles); $(\mhalo, \md)$ (rectangles); and $(\mhalo,\mh)$ (black dots).
} 
\label{massrelation} 
\bc
 \includegraphics[width=75mm]{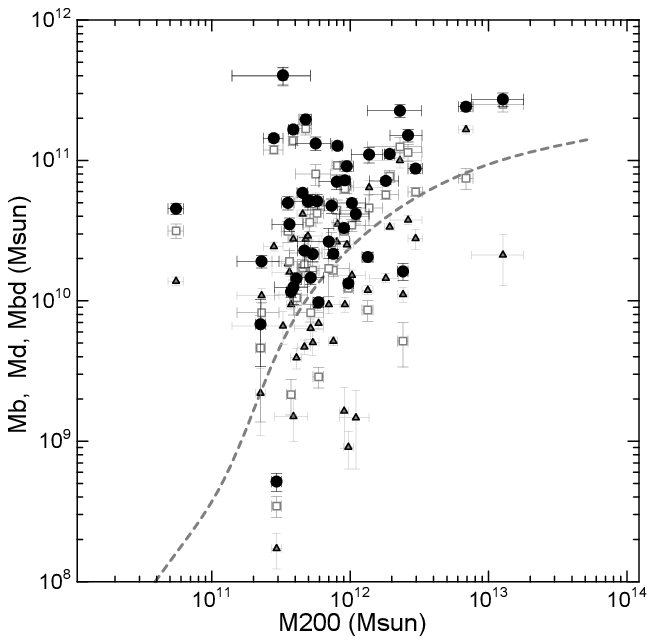} 
 \ec
\caption{Mass-Mass relations for $(\mhalo,\mb)$ (triangles); $(\mhalo,\md)$ (rectangles); and $(\mhalo,\mbd)$ (black dots). Dashed gray line shows the cosmological simulation + photometry (Behroozi et al. 2013). Halo mass $\mhalo$ may be multiplied by a factor of $\hub^2$ for different $H_0$ from 72 \kmsmpc.
} 
\label{massrelation_b} 
\end{figure}

The mass-mass plots are shown in figures \ref{massrelation} and \ref{massrelation_b}. As in the size-size relations, the more massive disks and halos are associated with more massive bulges and disks.  
As in table \ref{spearman} the mass-mass relation has higher correlation than size-size relation. The bulge and disk masses are correlated with coefficient as high as $\rhosp=0.63$. Moderate positive mass-mass correlations with $\rhosp \sim 0.33$ are found between bulge and halo and between disk and halo.  
 The tight correlation between $\mh$ and $\mhalo$ partially includes the trivial internal relation by equations (\ref{mh}) and (\ref{m200}). Note that the halo masses may be multiplied by a factor of $\hub^2$ to convert to values for different Hubble constant.

Figure \ref{massrelation_b} shows $\mbd$ plotted against $\mhalo$, which may be compared with the relation of stellar masses of galaxies against dark halo masses as obtained by cosmological simulation of star formation and hierarchical structure formation, as indicated by the gray dashed line for $z=0.1$ (Behroozi et al. (2013).

\begin{figure} 
\bc
 \includegraphics[width=75mm]{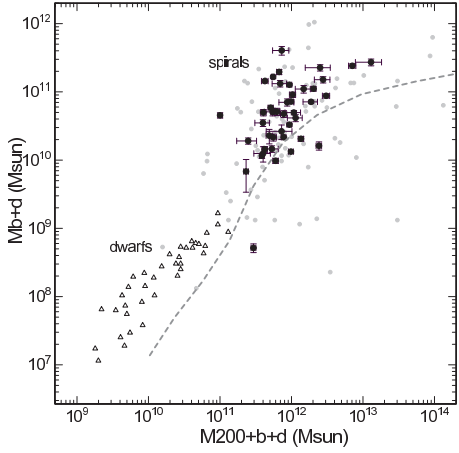} \ec
\caption{$\mbd$-$M_{\rm 200+b+d}$ relation compared with the stellar vs. total mass relation for dwarf galaxies (triangle: Miller et al. 2014) and simulation (gray dashed line: Behroozi et al. 2013). Black dots are the selected galaxies with reasonable fitting results, while small gray dots (including black dots) show non-weighted results from automatic decomposition of all rotation curves.} 
\label{mmdwsp} 
\end{figure}

Figure \ref{mmdwsp} shows the bulge+disk mass, $\mbd$, plotted against total mass, $M_{\rm 200+b+d}=\mhalo+\mbd$. Results for all the compiled rotation curves are shown by gray small dots. In the figure is also shown a plot of the photometric luminous mass against Virial mass for dwarf galaxies by Millar et al. (2014). These plots may be compared with a cosmological simulation combined with photometric stellar masses as shown by the dashed line (Behroozi et al. 2013).

 The simulation is in agreement in shape with the observations, while the simulated values of $\mbd$ are smaller on average by a factor of three than observed values. This may be partly due to the fact that the present masses are purely dynamical, so that the $\mbd$ values might contain dark matter mass. 

\subsection{Size-Mass relation}

\begin{figure} 
\bc
 \includegraphics[width=75mm]{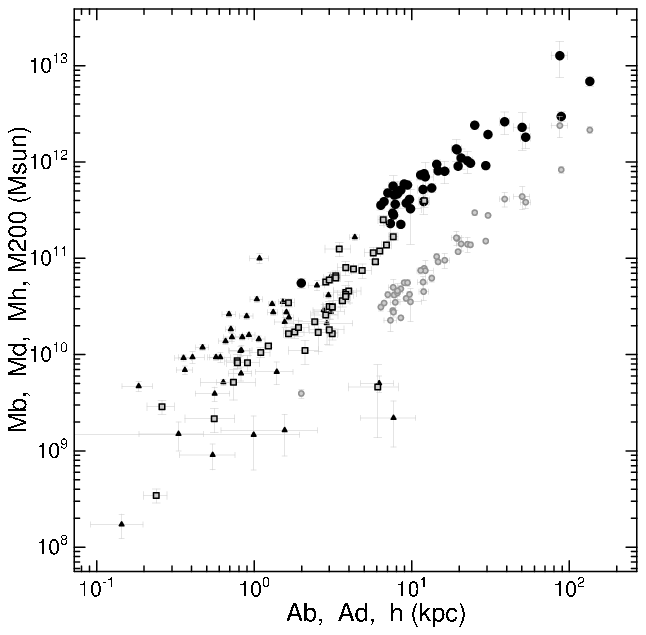} 
 \ec
\caption{Size-Mass relations for $(x,y)$ $=(\ab,\mb)$ (gray triangles); $(\ad,\md)$ (gray rectangles); $(h, \mh)$ (small dots); and $(h,\mhalo)$ (black dots).} 
\label{sizemass}
 
\bc
 \includegraphics[width=75mm]{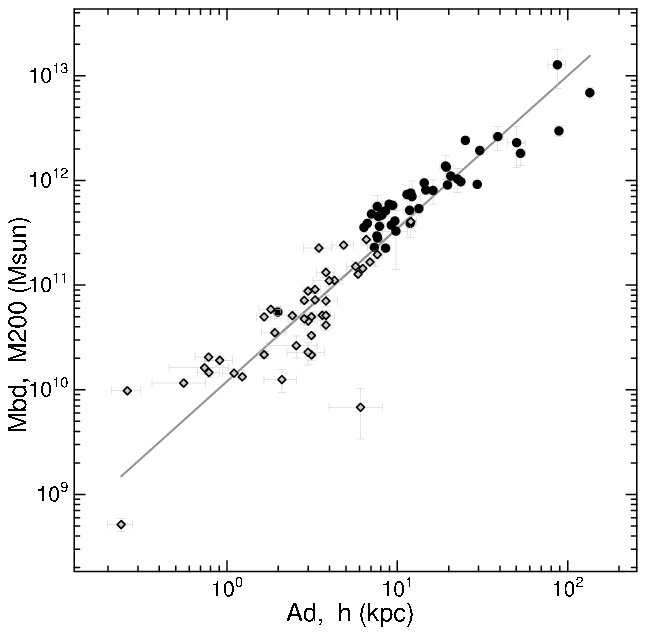} 
 \ec
\caption{Size-Mass relations for $(\ad,\mbd)$ (gray diamonds); and $(h,\mhalo)$ (black dots). } 
\label{sizemass_b} 
\end{figure}

\begin{figure}
\bc
 \includegraphics[width=75mm]{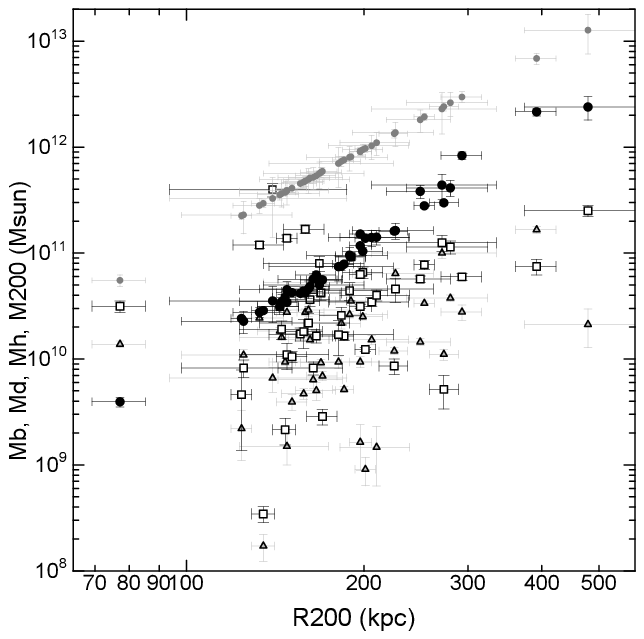} 
 \ec
\caption{Size-Mass relations for $(\rhalo,\mb)$ (triangles); $(\rhalo,\md)$ (rectangles); $(\rhalo,\mh)$ (black dots); $(\rhalo,\mhalo)$ (gray small dots) showing the trivial relation by equation \ref{m200}. } 
\label{sizemass_c} 
\end{figure}

The size-mass relations are shown in figure \ref{sizemass} to \ref{sizemass_c}. 
 The correlations between the size and mass for the bulge is as high as $r=0.87$ and $\rhosp=0.84$. 
It should be stressed that the positive size-mass relation applies to the dark halo.  
However, it should be remembered that the correlations are partially caused by the definition of the masses which include the scale radii in the forms as $M_{\rm b} \sim a_{\rm b}^2$, $M_{\rm d} \sim a_{\rm d}^2$, $M_{\rm h}, M_{200} \sim h^3$. 

The positive correlations between the size and mass both for bulges and disks, particularly the disk's size-mass relation, are the dynamical representation of the luminosity-size relation established by optical and infrared photometry (de Jong 1996; Graham and Worley 2008; Simard et al. 2011). .

The plotted size-mass relations can be represented by the following equations, which were obtained by the least-squares fitting in the log-log plane by linear functions. The mass and scale radii (in $\Msun$ and kpc, respectively) are given by 
\be
\log \mb =(10.06\pm 0.15)+( 0.72\pm 0.41) ~\log \ab ,
\label{logmbab}
\ee
\be
\log \md = (9.89\pm 0.23)+(1.38\pm 0.41) ~\log \ad ,
\label{logmdad}
\ee
and for the dark halo,
\be
\log \mh = (9.26\pm 0.52)+(1.45\pm 0.43) ~\log h.
\label{logmhh}
\ee 
 
The size-mass relation for the disk may be compared with the luminosity-size relation obtained by Simard et al. (2011) as
\be 
\mphoto_{\rm g, disk}\simeq -17.52 - 3.58 ~\log{ \ad },
\label{magsize}
\ee
or
\be 
\log L \simeq 8.97 +1.43 ~\log \ad ,
\label{lumisize}
\ee
where $\mphoto_{\rm g,disk}$ is the absolute magnitude in $g$-band of galaxy disks, and $L$ is the luminosity in $L_\odot$, and the coefficients were eye-estimated by fitting a straight line to the plot in their figure 7. 
Here derived equations \ref{logmdad} and \ref{lumisize} can be used to estimate an approximate mass-to-luminosity ratio of the disk in solar unit as
\be
\log(\md/L)\simeq 0.92 - 0.07 ~\log \ad.
\label{logmlratio}
\ee
Insertion of $a_{\rm d} \sim 3.3$ kpc yields  $\md/L\sim 7.7 M_\odot/L_\odot$. 
Here, $\md$ is the dynamical mass, and hence, it might contain dark matter in addition to stars and gas. 
 
 Figure \ref{sizemass_b} shows plots of $\mbd$ vs $\ad$ and $\mhalo$ vs $h$ in the same log-log plane. It is interesting that the two plots can be fitted by a single relation as indicated by the gray straight line, which represents a relation by the least-squares fitting,
\be
\log M_i= (10.18\pm 0.24) + (1.38\pm 0.21) \log a_i,
\label{fitmbdmhalo}
\ee
where $M_i=\mbd$ or $\mhalo$ in $\mten$ and $a_i=\ad$ or $h$ in kpc.
This simple equation leads to a relation between the bulge+disk mass to halo mass ratio expressed by the ratio of the scale radii of disk to halo as 
\be
\mbd/\mhalo\simeq (\ad/h)^{1.38}.
\ee
For the mean values of $\ad=3.3$ kpc and $h=21.6$ kpc, it leads to $\mbd/\mhalo \sim 0.07$.

Figure \ref{sizemass_c} shows the bulge, disk, and dark halo masses ($\mb, \md, \mh, \mhalo$) plotted against the critical halo radius $\rhalo$. The component masses are well positively correlated with the critical halo radius. The exact proportionality between the critical mass and radius, shown by gray small dots, is the trivial result of the relation $M_{200} \propto R_{200}^3$ defined by equation (\ref{m200}).

\subsection{Mass-to-Mass Ratio} 

\begin{figure}\bc
 \includegraphics[width=75mm]{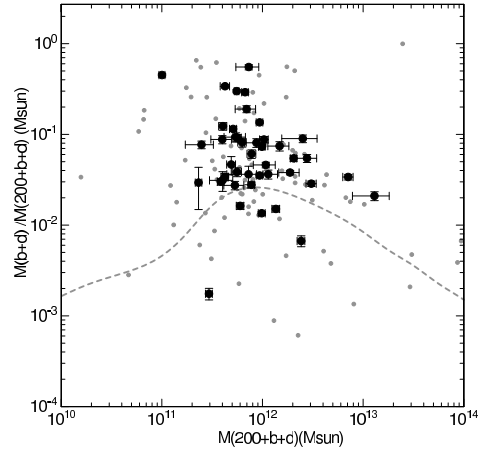} 
 \ec
\caption{Bulge + disk-to-total mass ratio plotted against total mass (black dots). Simulated relation by Behroozi et al. (2013) for $z=0.1$ is shown by a dashed gray line. Small gray dots show all the other galaxies not used in the analysis.} 
\label{massratio} 
\end{figure}

Figure \ref{massratio} shows the ratio of bulge + disk mass, $\mbd$, to the total mass , $M_{\rm 200+b+d}=\mbd+\mhalo$, plotted against the total mass, representing the bulge + disk mass fraction in the total mass. Results for all the other compiled galaxies are shown by gray small dots, which were, however, not used in the analyses.

Since $\mbd$ is approximately proportional to the luminosity $L$, the overall behavior in the figure indicates that the $M/L$ ratio decreases with luminosity. This is the well known relation from analyses of the universal rotation curves that $M/L$ ratio increases with decreasing luminosity, or less luminous galaxies are more dark matter dominant (Persic and Salucci 1995, 1996).

The mass ratios may be compared with cosmological simulations as indicated by a gray line in figure \ref{massratio} (Behroozi et al. 2013), who predicted baryon-to-dark matter ratio as small as $\sim 0.01$ for Milky Way sized galaxies.
This value is much smaller than the here obtained mean value of $\sim 0.06$ (table \ref{meanpara}) and the recent precise measurements for Galaxy and M31 of $\sim 0.07$ (Sofue 2015). Calibration of the models using the observed values in the local universe would be a subject for the future.

\subsection{Surface mass density}

\begin{figure*}\bc
 \includegraphics[width=150mm]{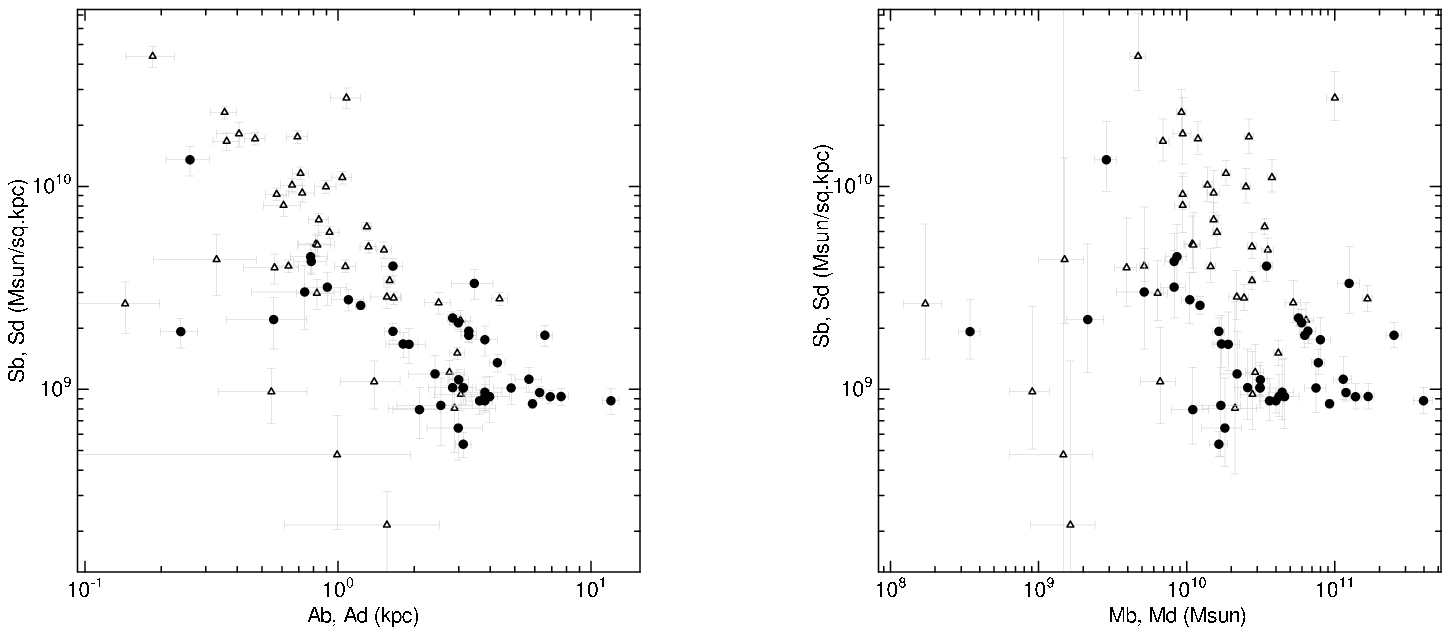} 
 \ec
\caption{Surface mass density (SMD) plotted against size and mass. Top panel: $(x,y)=(\ab,\sb)$ (triangles) and $(\ad,\sd)$ (black dots). Bottom: $(\mb,\sb)$ (triangles) and $(\md,\sd)$ (black dots). } 
\label{smd} 
\end{figure*}

Figure \ref{smd} plots the surface mass density (SMD) of bulges and disks defined by $S_{\rm b}=M_{\rm b}/\pi a_{\rm b}^2$ and $S_{\rm d}=M_{\rm d}/\pi a_{\rm d}^2$ against the scale radius and mass. The SMD decreases with the scale radius, which is partly due to inclusion of square of radius in the numerator of vertical axis. The inverse correlations are consistent with those between the scale radius and surface brightness observed by optical and near infrared photometry of face-on galaxies (de Jong 1996; Graham and Worley 2008). 
On the other hand, SMD appears to be not well correlated with masses of the bulge and disk. 

\section{Discussion}

\subsection{Summary}
 
\def\noi{\noindent}

The results may be summarized as follows:

\noi i) Nearly all published rotation curves for nearby disk galaxies in the last two decades were compiled from the literature .

\noi ii) Rotation curves were deconvolved into de Vaucouleurs bulge, exponential disk, and NFW dark halo by the simplified least $\chi^2$ fitting in order to determine the dynamical parameters (scale radii and masses) of individual components.  
 
\noi iii) Correlation analyses were obtained among the derived parameters, and empirical size-mass relations were derived for bulge, disk and dark halos. 
Correlation coefficients were found to be positive for all the plotted relations. Tight correlation was found between disk size and mass with $r=0.87$ and $\rhosp=0.84$. The bulge and disk masses are correlated with the halo mass at $r\sim 0.34$ by a chance occurence probability $\le 3$\%. 

\noi v) The bulge+disk mass to dark halo mass ratio of spiral galaxies is compared with that for dwarf galaxies, and the observed ratios are higher than that from simulations by a factor of three.
 
\noi vi) As a byproduct of the statistics, preliminary mass functions were derived, which were found to be represented by the Schechter function, with high mass end being better approximated by a power law.

\subsection{Mass Function of Dark Halos}

\def\F{\Phi}

The used rotation curve data cover most of typical spiral galaxies from various observations, and may be considered to be a representative sample of spirals in the nearest local universe at distances $\sim 20$ Mpc.
Although the sample is not large enough, preliminary size and mass functions may be constructed using the determined parameters for these galaxies.

 Denoting the parameters by $p_i$, which is equal either to $\ab,~\ad,~h,~\rhalo,~\mb,~\md,~\mh,$ or $~\mhalo$, the size and mass functions, $\F_i$ are defined by
 \be
 \F_i=dN/d {\rm log} p_i,
\label{smfunc}
\ee
where $N$ is the number of galaxies having parameter values from ${\rm log} p_i$ to ${\rm log} p_i+\delta {\rm log} p_i$. 

\begin{figure} 
\bc
\includegraphics[width=75mm]{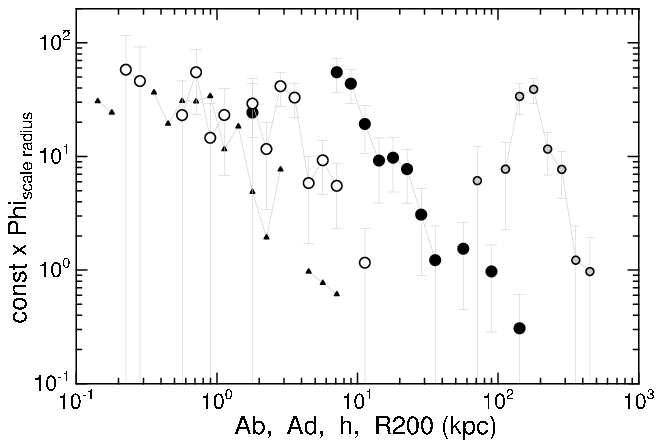} 
\ec
\caption{Size functions $\F_i$ for the bulge ($\ab$, triangle), disk ($\ad$, rectangle), dark halo scale radius ($h$, black dot), and dark halo critical radius ($\rhalo$, gray dot). Vertical axis is arbitrary, relative to the number of galaxies per unit dex interval of scale radius. } 
\label{sizefunc} 
\bc
\includegraphics[width=75mm]{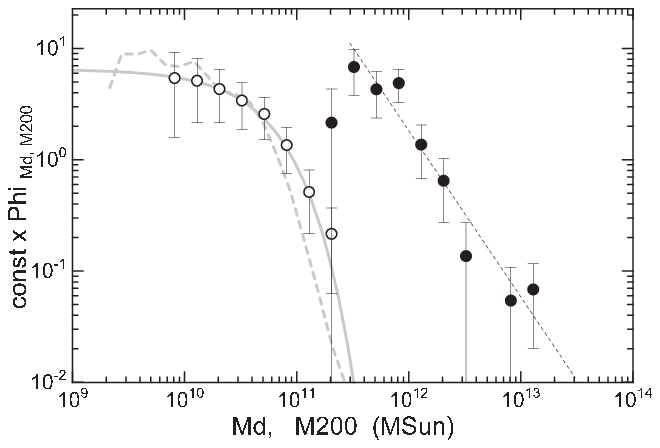} 
\ec
\caption {Mass functions $\F_i$ for the disk ($\md$, open circle) and dark halo ($\mhalo$, black dot). The gray dashed line shows a photometric result for late type galaxies (Bell et al. 2003), and the gray line is the Schechter function for $M_*=4\times 10^{10}\Msun$. The high-mass slope of dark halos is approximately by a power law of index $\sim -1.8$ as indicated by the dashed straight line. Vertical axis is arbitrary, proportional to the number of galaxies per dex interval of mass.}
\label{massfunc} 
\end{figure}

 Figures \ref{sizefunc} and \ref{massfunc} show the obtained size and mass functions, respectively, in relative units. Errors indicated by the bars were evaluated by the square root of the number of galaxies in each bin of counting at equal dex interval of the size and mass. The size and mass functions show similar behaviors, reflecting the size-mass relation. 

The shapes of the mass functions (figure \ref{massfunc}) for the bulge and disk are consistent with those obtained by photometric observations represented by the Schechter function (Schechter 1976; de Jong 1996; Bell et al 2003; 
Graham and Worley 2008; Cameron et al. 2009; Li \& White 2009 ;Simard et al. 2011; Lackner and Gunn 2012; Bernardi et al. 2013;
). 

In figure \ref{massfunc} a photometric mass function for late type galaxies as obtained by Bell et al. (2003) is shown by the thick dashed gray line. The observed distribution is well approximated by the Schechter function with $M_*=4 \times 10^{10}\Msun$, as indicated by the gray line. Here, the vertical values are so adjusted to fit the flat part. 

On the other hand, the dark halo mass function can be represented by a power law as indicated by the dashed line in figure \ref{massfunc}, 
\be
\F_{\mhalo} \sim dN/d {\rm log} \mhalo \propto \mhalo^{-1.8}.
\label{m200func}
\ee 
This may be compared with the calculated mass function by $N$-body numerical simulations (Brainerd and Villumsen 1992; Jenkins et al. 2001; Reed et al. 2003; Tinker et al. 2008), which predict power law behavior followed by Schechter-type steeper decrease at higher mass ends beyond $10^{14}\Msun$.

\subsection{Limitation and improvement by hybrid decomposition}
 
The sizes and masses of bulge, disk and dark halo were simultaneously determined for each galaxy, where the errors were evaluated as the ranges of the parameters around their best-fitting values that allow for 10\% increase of the least $\chi^2$ values. In this context and in so far as the selected 43 galaxies that satisfied the condition equation \ref{condition} are concerned, the result seems reliable. However, the disk-halo degeneracy problem (Bershardy et al. 2010) may still exist in many other galaxies.  
The dynamical decomposition, thus, requires rotation curves with sufficient accuracy in order to obtain realistic results. In fact, the selection criterion about the scale radii in order to avoid unrealistic fitting has caused rejection of a considerable number of galaxies from the compilation. 

It is also difficult to discriminate the physical compositions in the bulge, disk or dark halo, because the derived parameters are purely dynamical quantities. Namely, the bulge and disk masses might contain dark matter, and/or the dark halo mass might contain baryonic mass.
Photometric measurements of scale radii of bulge and disk would help to increase the number of sample galaxies, when rotation curves have poorer accuracy. Such hybrid decomposition will provide with a larger number of sample galaxies and a more reliable correlation analysis.

\begin{appendix}

\section{Atlas of Rotation Curves and Fitting Results}
 
All rotation curves are presented at URL: http://www.ioa.s.u-tokyo.ac.jp/~sofue/2015rc/ .

 \end{appendix}

\begin{thebibliography}{}
\def\r{\bibitem[]{}} 
\def\mnras{MNRAS}
\def\apj{ApJ}
\def\aap{AA}
\def\pasj{PASJ}
\def\apjs{ApJS}
\def\aj{AJ} 

\bibitem[Bershady et al.(2010)]{2010ApJ...716..198B} Bershady, M.~A., 
Verheijen, M.~A.~W., Swaters, R.~A., et al.\ 2010, \apj, 716, 198 

\bibitem[Behroozi et al.(2013)]{2013ApJ...770...57B} Behroozi, P.~S., 
Wechsler, R.~H., \& Conroy, C.\ 2013, \apj, 770, 57 

\bibitem[Bell et al.(2003)]{2003ApJS..149..289B} Bell, E.~F., McIntosh, 
D.~H., Katz, N., \& Weinberg, M.~D.\ 2003, \apjs, 149, 289 

\bibitem[Bernardi et al.(2013)]{2013MNRAS.436..697B} Bernardi, M., Meert, 
A., Sheth, R.~K., et al.\ 2013, \mnras, 436, 697 

\bibitem[\protect\citeauthoryear{Brainerd 
\& Villumsen}{1992}]{1992ApJ...394..409B} Brainerd T.~G., Villumsen J.~V., 1992, ApJ, 394, 409 
 
\bibitem[Cameron et al.(2009)]{2009ApJ...699..105C} Cameron, E., Driver, 
S.~P., Graham, A.~W., \& Liske, J.\ 2009, \apj, 699, 105 
 pc 
\bibitem[de Blok et al.(2008)]{2008AJ....136.2648D} de Blok, W.~J.~G., 
Walter, F., Brinks, E., et al.\ 2008, \aj, 136, 2648 

\bibitem[de Jong(1996)]{1996A&A...313...45D} de Jong, R.~S.\ 1996, \aap, 313, 45 
 
\bibitem[\dv(1958)]{1958ApJ...128..465D} \dv, G.\ 1958, \apj, 128, 465 

\bibitem[Dunkley et al.(2009)]{2009ApJ...701.1804D} Dunkley, J., Spergel, 
D.~N., Komatsu, E., et al.\ 2009, \apj, 701, 1804 

\bibitem[\protect\citeauthoryear{Erroz-Ferrer et al.}{2012}]{2012MNRAS.427.2938E} Erroz-Ferrer S., et al., 2012, MNRAS, 427, 
2938 


\r Freeman, K. C., 1970, ApJ, 160, 811

\bibitem[Garrido et al.(2005)]{2005MNRAS.362..127G} Garrido, O., Marcelin, 
M., Amram, P., et al.\ 2005, \mnras, 362, 127 

\bibitem[\protect\citeauthoryear{Gentile et al.}{2015}]{2015A&A...576A..57G} Gentile G., et al., 2015, A\&A, 576, A57 

\bibitem[\protect\citeauthoryear{Gentile et al.}{2007}]{2007MNRAS.375..199G} Gentile G., Salucci P., Klein U., Granato G.~L., 2007, MNRAS, 375, 199 

\bibitem[Graham 
\& Worley(2008)]{2008MNRAS.388.1708G} Graham, A.~W., \& Worley, C.~C.\ 2008, \mnras, 388, 1708 
 
\bibitem[Hinshaw et al.(2009)]{2009ApJS..180..225H} Hinshaw, G., Weiland, 
J.~L., Hill, R.~S., et al.\ 2009, \apjs, 180, 225 

\bibitem[\protect\citeauthoryear{Hlavacek-Larrondo et al.}{2011}]{2011MNRAS.416..509H} Hlavacek-Larrondo J., Marcelin M., Epinat B., Carignan C., de Denus-Baillargeon M.-M., Daigle O., Hernandez O., 2011a, 
MNRAS, 416, 509 

\bibitem[\protect\citeauthoryear{Hlavacek-Larrondo et 
al.}{2011}]{2011MNRAS.411...71H} Hlavacek-Larrondo J., Carignan C., Daigle 
O., de Denus-Baillargeon M.-M., Marcelin M., Epinat B., Hernandez O., 2011b, 
MNRAS, 411, 71 
 
\bibitem[\protect\citeauthoryear{Jenkins et 
al.}{2001}]{2001MNRAS.321..372J} Jenkins A., Frenk C.~S., White S.~D.~M., 
Colberg J.~M., Cole S., Evrard A.~E., Couchman H.~M.~P., Yoshida N., 2001, 
MNRAS, 321, 372 
 
\bibitem[Kent(1985)]{1985ApJS...59..115K} Kent, S.~M.\ 1985, \apjs, 59, 115 
 
\bibitem[Lackner 
\& Gunn(2012)]{2012MNRAS.421.2277L} Lackner, C.~N., \& Gunn, J.~E.\ 2012, \mnras, 421, 2277 

\bibitem[Li \& White(2009)]{2009MNRAS.398.2177L} Li, C., \& White, S.~D.~M.\ 2009, \mnras, 398, 2177 

\bibitem[\protect\citeauthoryear{M{\'a}rquez et al.}{2002}]{2002A&A...393..389M} M{\'a}rquez I., Masegosa J., Moles M., Varela J., Bettoni D., Galletta G., 2002, A\&A, 393, 389 

\bibitem[Martinsson et 
al.(2013)]{2013A&A...557A.131M} Martinsson, T.~P.~K., Verheijen, M.~A.~W., Westfall, K.~B., et al.\ 2013, \aap, 557, AA131 

\bibitem[\protect\citeauthoryear{Miller et al.}{2014}]{2014ApJ...782..115M} 
Miller S.~H., Ellis R.~S., Newman A.~B., Benson A., 2014, ApJ, 782, 115 

\bibitem[Moster et al.(2013)]{2013MNRAS.428.3121M} Moster, B.~P., Naab, T., 
\& White, S.~D.~M.\ 2013, \mnras, 428, 3121 

\r Nakanishi, H., Sofue, Y. 2006 PASJ 58, 847. 

\r Navarro, J. F., Frenk, C. S., White, S. D. M., 1996, ApJ, 462, 563 
 
\bibitem[Navarro et al.(1997)]{1997ApJ...490..493N} Navarro, J.~F., Frenk, 
C.~S., \& White, S.~D.~M.\ 1997, \apj, 490, 493 

\bibitem[Noordermeer et al.(2007)]{2007MNRAS.376.1513N} Noordermeer, E., 
van der Hulst, J.~M., Sancisi, R., Swaters, R.~S., 
\& van Albada, T.~S.\ 2007a, \mnras, 376, 1513 
 


\bibitem[\protect\citeauthoryear{Reed et al.}{2003}]{2003MNRAS.346..565R} 
Reed D., Gardner J., Quinn T., Stadel J., Fardal M., Lake G., Governato F., 
2003, MNRAS, 346, 565 


\bibitem[\protect\citeauthoryear{Reyes et al.}{2012}]{2012MNRAS.425.2610R} Reyes R., Mandelbaum R., Gunn J.~E., Nakajima R., Seljak U., Hirata C.~M., 2012, MNRAS, 425, 2610 



\bibitem[\protect\citeauthoryear{Olling}{1996}]{1996AJ....112..457O} Olling R.~P., 1996, AJ, 112, 457 
 
\r Rubin VC, Ford Jr WK, Thonnard N. 1980. ApJ 238:471 

\r Rubin VC, Ford Jr WK, Thonnard N. 1982. ApJ 261:439 

\r Persic M, Salucci P. 1995. Ap. J. Supp. 99:501

\r Persic M, Salucci P, Stel F. 1996. MNRAS 281:27 

\bibitem[\protect\citeauthoryear{Ryder et al.}{1998}]{1998MNRAS.293..411R} Ryder S.~D., Zasov A.~V., McIntyre V.~J., Walsh W., Sil'chenko O.~K., 1998, MNRAS, 293, 411 
 
\bibitem[Schechter(1976)]{1976ApJ...203..297S} Schechter, P.\ 1976, \apj, 
203, 297 

\bibitem[Simard et al.(2011)]{2011ApJS..196...11S} Simard, L., Mendel, 
J.~T., Patton, D.~R., Ellison, S.~L., 
\& McConnachie, A.~W.\ 2011, \apjs, 196, 11 


\bibitem[\protect\citeauthoryear{Sofue}{2015}]{2015PASJ..tmp..198S} Sofue 
Y., 2015, PASJ, 198 


\bibitem[\protect\citeauthoryear{Sofue}{2013}]{2013PASJ...65..118S} Sofue 
Y., 2013, PASJ, 65, 118 




\bibitem[\protect\citeauthoryear{Sofue et al.}{2003}]{2003PASJ...55...59S} 
Sofue Y., Koda J., Nakanishi H., Onodera S., 2003, PASJ, 55, 59 

\bibitem[Sofue et al.(1999)]{1999ApJ...523..136S} Sofue, Y., Tutui, Y., 
Honma, M., et al.\ 1999, \apj, 523, 136 

\bibitem[\protect\citeauthoryear{Sofue \& Rubin}{2001}]{2001ARA&A..39..137S} Sofue Y., Rubin V., 2001, ARA\&A, 39, 137 



\bibitem[Swaters et al.(2009)]{2009A&A...493..871S} Swaters, R.~A., Sancisi, R., van Albada, T.~S., \& van der Hulst, J.~M.\ 2009, \aap, 493, 871 
\bibitem[\protect\citeauthoryear{Tinker et al.}{2008}]{2008ApJ...688..709T} 
Tinker J., Kravtsov A.~V., Klypin A., Abazajian K., Warren M., Yepes G., 
Gottl{\"o}ber S., Holz D.~E., 2008, ApJ, 688, 709 


\bibitem[\protect\citeauthoryear{Whitmore, McElroy, \& Schweizer}{1987}]{1987ApJ...314..439W} Whitmore B.~C., McElroy D.~B., Schweizer F., 1987, ApJ, 314, 439 

\bibitem[Yoshino 
\& Ichikawa(2008)]{2008PASJ...60..493Y} Yoshino, A., \& Ichikawa, T.\ 2008, \pasj, 60, 493 
 


\end{thebibliography}
\end{document}